\documentclass{article}
\usepackage{spconf,hyperref}
\usepackage{tikz}
\usepackage{amsmath}

\usepackage{algorithmic}
\usepackage{xcolor}
\usepackage{xspace}
\usepackage{tcolorbox}
\usepackage{soul}
\usepackage{booktabs}
\usepackage{threeparttable}
\usepackage{pifont}
\usepackage{tabularx}
\usepackage{multirow}
\usepackage{graphicx}%插入图片
\usepackage{subfigure} %子图
\usepackage[linesnumbered,ruled,vlined]{algorithm2e}
\usepackage{bbm}
\usepackage{amsmath}
\usepackage{amsfonts,amssymb}
\usepackage{float}
\usepackage{bm}
\usepackage{url}

\newcommand{\sys}{{\sc Synaspot}\xspace}

\title{Synaspot: A Lightweight, Streaming Multi-modal Framework for Keyword Spotting with Audio-Text Synergy}
\name{Kewei Li$^\text{\textdagger}$, Yinan Zhong$^\text{\textdagger}$, Xiaotao Liang$^{*}$, Tianchi Dai, Shaofei Xue\thanks{$^\text{\textdagger}$ Equal contribution.  $^*$ Corresponding author.}}
\address{Intelligent Connectivity, Alibaba Group, Hangzhou, China\\\{weike.lkw, zhongyinan.zyn, xiaotao.lxt, daitianchi.dtc, shaofei.xsf\}@alibaba-inc.com}

\begin{document}
%\ninept

\maketitle
\begin{abstract}

Open-vocabulary keyword spotting (KWS) in continuous speech streams holds significant practical value across a wide range of real-world applications.
While increasing attention has been paid to the role of different modalities in KWS, their effectiveness has been acknowledged. However, the increased parameter cost from multimodal integration and the constraints of end-to-end deployment have limited the practical applicability of such models. To address these challenges, we propose a lightweight, streaming multi-modal framework.
First, we focus on multimodal enrollment features and reduce speaker-specific (voiceprint) information in the speech enrollment to extract speaker-irrelevant characteristics. Second, we effectively fuse speech and text features. Finally, we introduce a streaming decoding framework that only requires the encoder to extract features, which are then mathematically decoded with our three modal representations.
Experiments on LibriPhase and WenetPrase demonstrate the performance of our model. Compared to existing streaming approaches, our method achieves better performance with significantly fewer parameters.

% 当越来越多唤醒关注在不同的的模态作用上，我们关注到了它的有效性，但是多模态带来的参数增加和端到端的部署降低了模型的实用性。因此，我们提出了一个 LIGHTWEIGHT, STREAMING MULITI-MODAL FRAMEWORK。首先，我们关注到多模态的注册特征，对speech的注册进行说话人的speaker-irrelevant To reduce speaker-specific (voiceprint) information，其次，我们将speech特征与text特征进行融合，最后，我们提出了一套流式的解码框架，只需encoder提取特征与我们的三个模态的特征做数学解码。实验在LibriPrase上验证了我们模型的性能，与流式的方法相比，我们能达到更少的参数量同时实现更好的性能。

\end{abstract}
\begin{keywords}
Keyword spotting, open vocabulary, contrastive learning, streaming.
\end{keywords}

\section{Introduction}\label{sec:Introduction}

Keyword spotting (KWS) is a technology designed to continuously detect the presence of keywords in audio streams, it is now serving as a critical entrance to human-machine interaction systems. Since KWS is typically deployed on resource-constrained devices, models must be compact and low-latency, which has driven the development of small-footprint, streaming KWS.

KWS systems broadly fall into two categories. The first is customized KWS (e.g., “Ok Google”), which is widely used for device wake-up. The other one is open-vocabulary KWS, it is emerged to improve user flexibility, allowing users to specify keywords on demand. Our work focuses on the latter. 
Existing open-vocabulary KWS approaches include audio enrollment methods such as Dynamic Time Warping (DTW) \cite{Fuchs2017SpokenTD, Yusuf2019AnEE} and metric learning \cite{MultilingualqbeReuter}, as well as text enrollment \cite{ai2024mm, ContrastiveXi, plcl}. 
It has been shown that multi-modal enrollment features can increase both the expressiveness and accuracy of KWS models. In particular, cross-modal alignment features enhance performance overall, and treating audio alignment as auxiliary supervision offers further improvements \cite{ai2024mm}. Combining several aligned modalities delivers additional representational benefits \cite{plcl}.
Standing on these insights, we equip the proposed framework, \sys, with multi-modal enrollment (audio, text, and mixed), offering users a flexible choice of enrollment modes and enabling robust operation across diverse scenarios. For audio-only enrollment, we develop an audio encoder and enhance it with speaker domain adaptation. For text-only enrollment, we apply contrastive learning (CL) to align text feature space. When both modalities are available, we further incorporate a fused feature that combines audio and text modalities, increasing enrollment diversity and enhancing KWS accuracy.

To compare enrolled keywords with user-spoken speech, many existing decision methods rely on non-streaming end-to-end models, such as GRU combined with fully connected layers, for discrimination\cite{nishu2024flexible,lee2024iphonmatchnet,Shin2022LearningAA}. However, such models require audio segments with precisely aligned word boundaries, which makes them unsuitable for real-world streaming scenarios. However, many lightweight streaming models\cite{xi2025streaming, garai2025advances} are limited to customized KWS, lacking open-vocabulary capability.
To address this limitation, window-based streaming approaches have been proposed, where end-to-end decisions are performed over fixed-length windows\cite{DohyunKim}. Yet, this design leads to significant performance variations when the lengths of keywords differ.
To overcome these issues, we propose a decoding framework that leverages only encoder representations and enrollment features for mathematical decoding. This enables the recognition of enrolled keyword with varying lengths directly from streaming audio, while maintaining a lightweight model that consists of a single encoder.
Finally, we introduce a complete architecture that integrates multimodal enrollment features with streaming audio decoding. This design achieves a lightweight KWS system with both low latency and compact model size.

In summary, our contributions are concluded as follows:

\begin{itemize}
    \item We propose \sys, a lightweight, streaming multi-modal framework for KWS, fusing three modalities (audio, text, audio-text mixed).
    \item We develop a multimodal flexible enrollment keyword spotting system.
    \item We conduct series of experiments to validate the effectiveness of \sys. The results demonstrate that \sys outperforms baselines in both English and Chinese KWS tasks.
\end{itemize}

\begin{figure*}[tt]
    \centering
\setlength{\abovecaptionskip}{0cm}
\setlength{\belowcaptionskip}{0cm}

\centerline{\includegraphics[width=\linewidth]{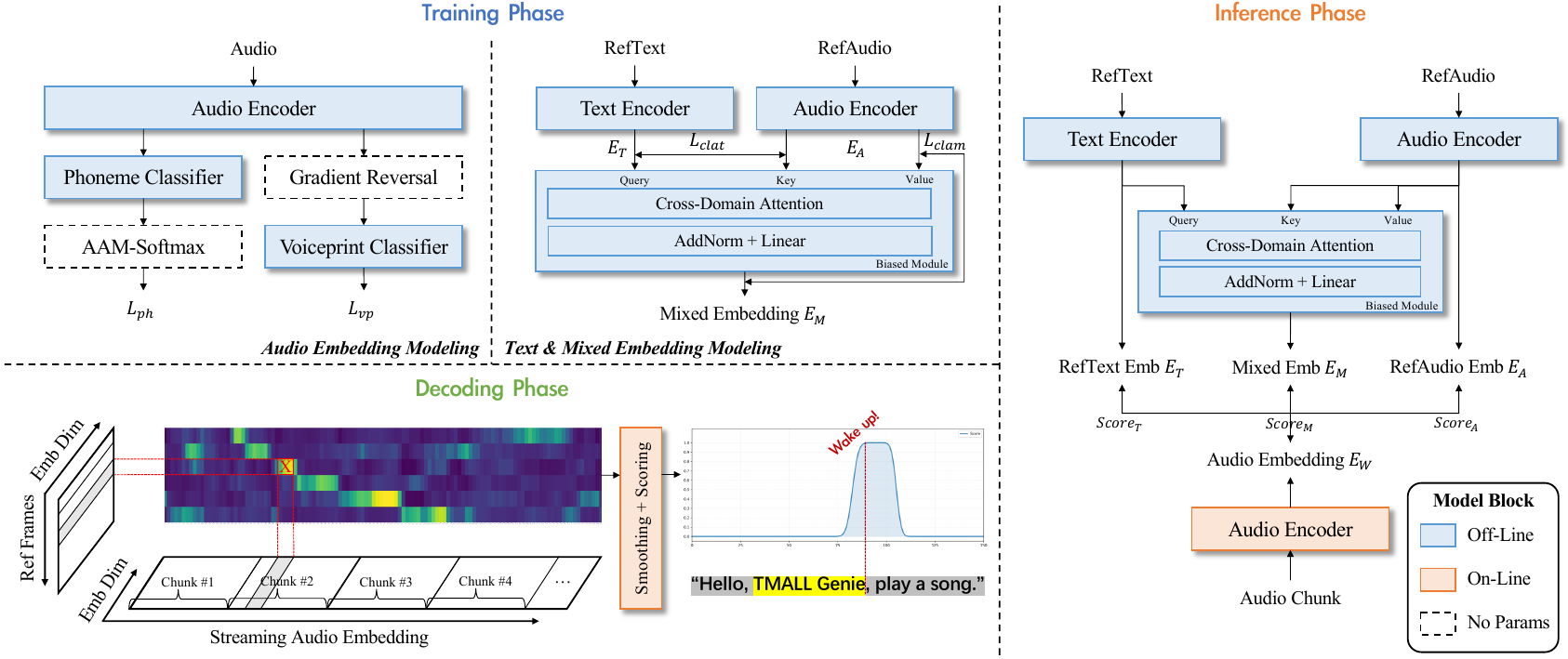}}

\caption{An overview of \sys. In \ding{192} the training phase, we first learn a speaker-irrelevant audio encoder and explicitly enlarge the inter-phoneme margins, then we obtain text \& mixed embeddings and align these modalities in a shared embedding space via contrastive learning. In \ding{193} the inference \& decoding phase, we perform a streaming and lightweight keyword spotting, assign each audio frame with scores.}\label{fig:method}
\end{figure*}
% \input{Tables/algorithm}
% % \input{Figures/heatmap}
% \clearpage

\section{\sys: Design Details}\label{sec:method}

As shown is Fig.~\ref{fig:method}, \sys consists of two phases, i.e., training phase, and inference \& decoding phase.

\subsection{Training Phase}
\subsubsection{Audio Embedding Modeling}\label{subsubsec:audio_modeling}

In the training phase, we first learn a speaker-irrelevant audio encoder and explicitly enlarge the inter-phoneme margins.

The audio encoder comprises $L$ DFSMN~\cite{zhang2018deep} layers. It takes audio FBank frames as input and outputs frame-level audio embeddings $E_A$. A phoneme classifier is then applied to $E_A$ to predict the label of each frame. To reduce phoneme confusion and false positives, we adopt the additive angular margin (AAM) loss~\cite{desplanques2020ecapa}. The objective is defined in Function~\ref{equ:aamloss}, where $\theta_{j}$ represents the angle between the model weight vector $\mathbf{w}_j$ and the input $x_i$, with~$s$ serving as the rescaling parameter, and~$m$ acting as the angular margin~penalty.

\begin{equation}\label{equ:aamloss}
\mathcal{L}_\mathrm{ph}= -\frac1N\sum_{i=1}^N\log\frac{e^{s(\cos(\theta_{y_i}+q))}}{e^{s(\cos(\theta_{y_i}+q))}+\sum_{j=1,j\neq y_i}^le^{s(\cos(\theta_{j}))}}
\end{equation}

To reduce speaker-specific (voiceprint) information, we apply domain adaptation~\cite{ganin2015unsupervised} to \sys by attaching a speaker classifier to the audio encoder and reversing its gradients during training. This encourages utterances with the same linguistic content but different speakers to produce similar embeddings, thereby improving the effectiveness of enrollment audio in the keyword spotting task. The speaker classifier consists of an attentive pooling layer~\cite{desplanques2020ecapa} followed by a linear classifier and is optimized using the cross-entropy loss, denoted as $\mathcal{L}_{vp}$.

We jointly optimize the two objectives as shown in Function~\ref{equ:audioloss}, where $\alpha_A$ and $\beta_A$ are hyper-parameters that control their relative contributions.

\begin{equation}\label{equ:audioloss}
\mathcal{L}_\mathrm{audio}= \alpha_A \cdot \mathcal{L}_\mathrm{ph} + \beta_A \cdot \mathcal{L}_\mathrm{vp}
\end{equation}

\subsubsection{Text \& Mixed Embedding Modeling}

We obtain text embeddings $E_T$ using an embedding layer followed by an LSTM~\cite{hochreiter1997long}. To form the mixed embedding $E_M$, we apply a cross-attention layer in which $E_T$ serves as the queries and $E_A$  serves as the keys and values. We then align the audio, text and mixed modalities in a shared embedding space via contrastive learning~\cite{ContrastiveXi}. The objectives are defined in Function~\ref{equ:clatloss} and Function~\ref{equ:clamloss}, where $sim$ denotes the similarity function (cosine similarity on L2-normalized embeddings unless otherwise noted), $E_*^i$ means the $i^{th}$ frame of embedding $E_*$, $\tau \textgreater 0$ is the temperature, and $i \neq j$.

\begin{equation}\label{equ:clatloss}
\mathcal{L}_{\mathrm{clat}}=-\sum_{i=1}^N\log\frac{\exp(sim({E}_T^i,{E}_A^i)/\tau)}{\sum_{j=1}^N\exp(sim({E}_T^i,{E}_A^j)/\tau)}
\end{equation}

\begin{equation}\label{equ:clamloss}
\mathcal{L}_{\mathrm{clam}}=-\sum_{i=1}^N\log\frac{\exp(sim({E}_M^i,{E}_A^i)/\tau)}{\sum_{j=1}^N\exp(sim({E}_M^i,{E}_A^j)/\tau)}
\end{equation}

We use the audio encoder trained in Section~\ref{subsubsec:audio_modeling} and jointly optimize the three objectives as shown in Function~\ref{equ:mixedloss}, where $\alpha_M$, $\beta_M$ and $\gamma_M$ are hyper-parameters that control their relative contributions.

\begin{equation}\label{equ:mixedloss}
\mathcal{L}_\mathrm{mixed}= \alpha_M \cdot \mathcal{L}_\mathrm{ph} + \beta_M \cdot \mathcal{L}_\mathrm{clat} + \gamma_M \cdot \mathcal{L}_\mathrm{clam}
\end{equation}

\subsection{Inference \& Decoding Phase}

To enable streaming and lightweight keyword spotting, \sys first computes and caches three enrollment embeddings from the enrollment audio and text, i.e., the audio embedding $E_A$, the text embedding $E_T$, and the mixed audio–text embedding $E_M$. Once verified, these embeddings are fixed and need not be recomputed. Consequently, during online inference the only model executed is the audio encoder. \sys processes the incoming waveform in small, contiguous audio chunks, computes frame-level audio embeddings $E_W$ on the fly, and forwards them to the decoder to produce per-frame keyword scores.

In the decoding phase, we compute framewise similarity scores $P$ between the streaming audio embeddings $E_W$ and each enrollment embedding $E_\mathrm{Enroll} \in \{E_A, E_T, E_M\}$. Let $p_{ij}$ denote the similarity between $E_\mathrm{Enroll}^i$ and $E_W^j$.

Since raw scores can be noisy, we apply causal smoothing~\cite{Guoguochen} using a moving average:
\begin{equation}\label{equ:smooth}
p_{ij}^{\prime}=\frac{1}{j-h_\mathrm{smooth}+1}\sum_{k=h_\mathrm{smooth}}^jp_{ik}
\end{equation}
where $h_\mathrm{smooth} = max\{1,j-w_\mathrm{smooth} + 1\}$ is the index of the first frame within the smoothing window of size $w_\mathrm{smooth}$.

We then compute a confidence score at the $j^{th}$ frame by aggregating the best per-unit similarities within a causal scoring window:
\begin{equation}\label{equ:score}
Score=\sqrt[n-1]{\prod_{i=1}^{n-1}\max_{h_\mathrm{scoring}\leq k\leq j}p_{ik}^{\prime}}
\end{equation}
where $h_\mathrm{scoring} = max\{1,j-w_\mathrm{scoring} + 1\}$ is the index of the first frame within the scoring window of size $w_\mathrm{scoring}$. 

Depending on the enrollment modality, we obtain $Score_A$ (audio-only), $Score_T$ (text-only), or $Score_{All}=\alpha_S \cdot Score_A + \beta_S \cdot Score_T + \gamma_S \cdot Score_M$ (audio + text + mixed), where $\alpha_S$, $\beta_S$ and $\gamma_S$ are fusion weights.

% \begin{itemize}
%     \item audio-only: $Score_A$,
%     \item text-only: $Score_T$,
%     \item fused (audio + text + mixed): $Score_{All} = \alpha_S \cdot Score_A + \beta_S \cdot Score_T + \gamma_S \cdot Score_M$, where $\alpha_S$, $\beta_S$ and $\gamma_S$ are fusion weights.
% \end{itemize}

\begin{table}[tt]

\caption{Overall Effectiveness.}\label{tab:overall_effectiveness}

\small
\begin{center}
\begin{threeparttable}

\begin{tabularx}{\linewidth}{@{}l|c|c|>{\centering\arraybackslash}X>{\centering\arraybackslash}X|>{\centering\arraybackslash}X>{\centering\arraybackslash}X@{}}
% \begin{tabular}{c|cccccccccccccccc}
\toprule
\multicolumn{1}{c|}{\multirow{2}{*}{\textbf{Model}}} & \multicolumn{1}{c|}{\multirow{2}{*}{\textbf{EM}\tnote{\textdagger}}} & \multirow{2}{*}{\textbf{Params}} & \multicolumn{2}{c|}{\textbf{EER(\%)}} & \multicolumn{2}{c}{\textbf{AUC(\%)}} \\
\multicolumn{1}{c|}{}                                & \multicolumn{1}{c|}{}                             &                                   & $\textbf{LP}_{\textbf{E}}$\tnote{\textdaggerdbl}       & $\textbf{LP}_{\textbf{H}}$\tnote{\textdaggerdbl}       & $\textbf{LP}_{\textbf{E}}$       & $\textbf{LP}_{\textbf{H}}$      \\ \midrule
CMCD    \cite{Shin2022LearningAA}                                             & T                                                 & 0.7M                                  & 8.42              & 32.90             & 96.70             & 73.58            \\
Triplet     \cite{Triplet} & T                                                 & 0.6M                                 & 32.75             & 44.36             & 63.53             & 54.88            \\
SoftTripe    \cite{Huang2021QueryByExampleKS}                                        & T                                                 & N/A                                 & 28.74             & 41.95             & 78.74             & 62.65            \\
InfoNCE  \cite{ContrastiveXi}                                            & T                                                 & 2.2M                                 & 8.99              & 32.51             & 96.85             & 74.87            \\
CLAD    \cite{ContrastiveXi}                                         & T                                                 & 2.2M                             & 8.65              & 30.30             & 97.03             & 76.15            \\ \midrule
Synaspot-AA\tnote{\P} & A & 0.9M & 8.85     & 32.14             & 96.15    & 73.75            \\
Synaspot-AT\tnote{\P} & T & 0.9M                            & 7.07              & 28.69             & 97.17             & 77.35            \\
Synaspot                                             & TA                                                & 0.9M                            & \textbf{5.77}              & \textbf{27.29}    & \textbf{97.34}             & \textbf{79.15}   \\ 

 \bottomrule

% \end{tabular}
\end{tabularx}

\begin{tablenotes}[flushleft]
\footnotesize
\item[] \textdagger: Enrollment modality, T represents text, A represents audio.
\item[] \textdaggerdbl: $\mathrm{LP}_\mathrm{E}$, $\mathrm{LP}_\mathrm{H}$ represent Libriphase Easy and Hard, respectively.
\item[] \P: A represents audio enrollment only, T represents text enrollment only
\end{tablenotes}

\end{threeparttable}

\end{center}

\end{table}
\begin{table}[tt]

\caption{Chinese KWS Effectiveness.}\label{tab:effectiveness_cn}

\small
\begin{center}
\begin{threeparttable}

\begin{tabularx}{\linewidth}{@{}l|c|c|>{\centering\arraybackslash}X>{\centering\arraybackslash}X|>{\centering\arraybackslash}X>{\centering\arraybackslash}X@{}}
% \begin{tabular}{c|cccccccccccccccc}
\toprule
\multicolumn{1}{c|}{\multirow{2}{*}{\textbf{Model}}} & \multicolumn{1}{c|}{\multirow{2}{*}{\textbf{EM}\tnote{\textdagger}}} & \multirow{2}{*}{\textbf{Params}} & \multicolumn{2}{c|}{\textbf{EER(\%)}} & \multicolumn{2}{c}{\textbf{AUC(\%)}} \\
\multicolumn{1}{c|}{}                                & \multicolumn{1}{c|}{}                             &                                   & $\textbf{WP}_{\textbf{E}}$\tnote{\textdaggerdbl}       & $\textbf{WP}_{\textbf{H}}$\tnote{\textdaggerdbl}       & $\textbf{WP}_{\textbf{E}}$       & $\textbf{WP}_{\textbf{H}}$      \\ \midrule
MM-KWS\cite{ai2024mm} & T & 3.9M & 4.24 & 26.23 & 99.19 & 79.24            \\
MM-KWS-1.5s & T & 3.9M & 11.51 & 31.80 & 95.62 & 73.54            \\
MM-KWS-2.0s & T & 3.9M & 18.38 & 39.15 & 89.67 & 66.45            \\ \midrule
Synaspot-AA\tnote{\P} & A & 0.9M & 18.46 & 36.68 & 89.58    & 68.19 \\
Synaspot-AT\tnote{\P} & T & 0.9M & 19.76 & 39.95 & 87.25 & 64.70 \\
Synaspot & TA & 0.9M & 14.56 & 34.50 & 92.87 & 70.35   \\ 

\bottomrule

% \end{tabular}
\end{tabularx}

\begin{tablenotes}[flushleft]
\footnotesize
\item[] \textdagger: Enrollment modality, T represents text, A represents audio.
\item[] \textdaggerdbl: $\mathrm{WP}_\mathrm{E}$, $\mathrm{WP}_\mathrm{H}$ represent WenetPhrase Easy and Hard, respectively.
\item[] \P: A represents audio enrollment only, T represents text enrollment only
\end{tablenotes}

\end{threeparttable}

\end{center}

\end{table}
\begin{table}[tt]

\caption{Ablation Study.}\label{tab:ablation_study}

\small
\begin{center}
\begin{threeparttable}

\begin{tabularx}{\linewidth}{@{}l|>{\centering\arraybackslash}X>{\centering\arraybackslash}X|>{\centering\arraybackslash}X>{\centering\arraybackslash}X@{}}
% \begin{tabular}{c|cccccccccccccccc}
\toprule
\multicolumn{1}{c|}{\multirow{2}{*}{\textbf{Ablation}}} & \multicolumn{2}{c|}{\textbf{EER(\%)}} & \multicolumn{2}{c}{\textbf{AUC(\%)}} \\
\multicolumn{1}{c|}{}                                   & $\textbf{LP}_{\textbf{E}}$\tnote{\textdaggerdbl}       & $\textbf{LP}_{\textbf{H}}$\tnote{\textdaggerdbl}       & $\textbf{LP}_{\textbf{E}}$       & $\textbf{LP}_{\textbf{H}}$      \\ \midrule
% None\tnote{\textdagger} & 8.85     & 32.23             & 96.05    & 73.69     \\
% Synaspot & 8.85     & 32.23             & 96.05    & 73.69     \\
Synaspot & \textbf{5.77}              & \textbf{27.29}    & \textbf{97.34}             & \textbf{79.15}        \\
% w/o Mixed Embedding& 6.03             & 28.53    & 97.10             & 78.43             \\

w/o  Mixed Embedding & 7.07              & 29.09             & 97.04             & 76.85      \\
w/o Speaker Classifier & 8.85 & 32.04 & 95.87 & 72.90            \\ 

\bottomrule

% \end{tabular}
\end{tabularx}

\begin{tablenotes}[flushleft]
\footnotesize
% \item[] \textdagger: No ablation.
\item[] \textdaggerdbl: $\mathrm{LP}_\mathrm{E}$, $\mathrm{LP}_\mathrm{H}$ represent Libriphase Easy and Hard, respectively.
\end{tablenotes}

\end{threeparttable}

\end{center}

\end{table}

\section{Experimental}\label{sec:Experimental}

\subsection{Datasets}
The LibriSpeech \cite{LibrispeechPanayotov} train-clean-100 and train-clean-360 subsets are used for training. In addition, we employ the LibriPhrase \cite{Shin2022LearningAA} dataset, which consists of phrases extracted from LibriSpeech.
The evaluation dataset originates from the train-others-500 subset. Negative samples are divided into two categories: easy ($\textbf{LP}_\textbf{E}$) and hard ($\textbf{LP}_\textbf{H}$) based on their Levenshtein distance \cite{Levenshtein1965BinaryCC}, where the hard set ($\textbf{LP}_\textbf{H}$) contains a large number of phonetically or orthographically similar words.
% ai2024mm

For Mandarin validation, we employed the WenetPhrase dataset\cite{ai2024mm} containing two specialized subsets: WenetPhrase-Easy ($\textbf{WP}_\textbf{E}$) and WenetPhrase-Hard ($\textbf{WP}_\textbf{H}$). 
To simulate real-world streaming scenarios, we constructed the test set for decoding evaluation by randomly concatenating speech segments as acoustic buffers preceding and following both $\textbf{WP}_\textbf{E}$ and $\textbf{WP}_\textbf{H}$ utterances.

\subsection{Setup}
We implement a prototype of \sys in PyTorch.
% ~\cite{DBLP:conf/nips/PaszkeGMLBCKLGA19}. 
The audio encoder uses $L=7$ DFSMN layers, each has a hidden dimension of 256. We train for 50 epochs on eight NVIDIA RTX 3090 GPUs using Adam with an initial learning rate of 1e-3, an annealing factor of 0.3, and a batch size of 100.

During multimodal training, we weight the losses in Function(~\ref{equ:audioloss}) and (\ref{equ:mixedloss}) with $(\alpha_A,\beta_A)=(0.5,0.5)$ while $(\alpha_M,\beta_M,\gamma_M) 
= (0.4,0.3,0.3)$. At inference time, score-level fusion uses $(\alpha_S,\beta_S,\gamma_S) = (0.5,0.25,0.25)$.

\subsection{Overall Effectiveness}

We present the effectiveness of our streaming decoding approach in Table \ref{tab:overall_effectiveness}, evaluated using the AUC and EER metrics. To demonstrate streaming performance, each audio sample in the LibriPhrase dataset was randomly padded with additional audio segments both before and after. 
% The results show that our Synapot model, with only 0.9M parameters,  achieves an EER of 27.71\% and an AUC of 78.70\% on $\text{LP}_\text{H}$, and an EER of 5.97\% and an AUC of 97.19\% on $\text{LP}_\text{E}$. 
The results demonstrate that our Synaspot model achieves superior efficiency-performance balance, attaining 27.71\% EER and 78.70\% AUC on $\text{LP}_\text{H}$ along with 5.97\% EER and 97.19\% AUC on 
$\text{LP}_\text{E}$ under streaming conditions, while maintaining a compact parameter footprint of 0.9M.

% These outcomes confirm the feasibility of our streaming decoding method, highlight the benefits of incorporating multimodal information, and validate the proposed lightweight multimodal architecture.

We validated the feasibility of our streaming algorithm on the Mandarin WenetPhrase dataset.
As shown in Table. \ref{tab:effectiveness_cn}, our results in Mandarin demonstrate the generalizability of our method.

\subsection{Qualitative Analysis}
\begin{figure}[tt]
    \centering
\setlength{\abovecaptionskip}{0cm}
\setlength{\belowcaptionskip}{0cm}

\centerline{\includegraphics[width=7.6cm]{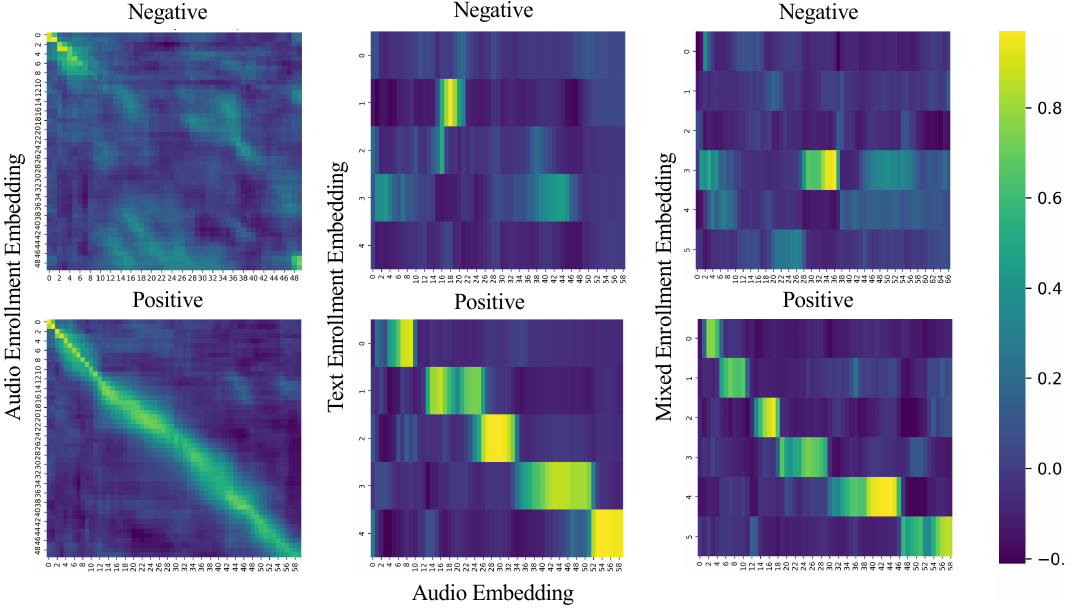}}
% \linewidth
\caption{Visualization of similarity heatmaps.}\label{fig:heatmap}
\end{figure}

We visualize the similarity heatmaps in Fig.~\ref{fig:heatmap}, which are derived from the similarity matrices of both offline embeddings (from audio, text, and mixed modalities) and online encoder outputs. These heatmaps clearly illustrate distinct highlighted regions between positive and negative examples, demonstrating the feasibility of our streaming decoding approach.
% We perform streaming decoding on these similarity maps, a
As illustrated in Fig.~\ref{fig:method}. By leveraging the pre-registered horizontal coordinates of the three modalities for scoring, a final wake-up score is computed, resulting from the collaborative integration of all three modalities.

% 有无mix模态
% 多speaker注册

\subsection{Ablation Study}

% Our ablation study results in Table \ref{tab:ablation_study} demonstrate that integrating both Mix Embedding and Speaker Classifier components enhances Synaspot's performance. This improvement validates the effectiveness of the combined feature representation. Specifically, the model achieves effective speaker disentanglement through discriminative characteristic separation while maintaining consistent robustness against enrollment data variations.

Our ablation study results in Table \ref{tab:ablation_study} demonstrates that combining Mix Embedding and Speaker Classifiers improves Synaspot performance, confirming the effectiveness of the integrated feature representation. The architecture achieves speaker-discriminative separation while maintaining robustness to enrollment data variations.

% \ref{tab:ablation_study}
% The experimental results, summarized in Table 3, reveal a substantial improvement in system performance attributable to the proposed Mix Embedding module and Speaker Classifier. This superior performance confirms that the integrated feature set promotes enhanced results. The model successfully achieves speaker disentanglement, distinguishing characteristics across different speakers, and maintains robust performance when faced with variability in the enrollment data.

\subsection{stream to end-to-end}
% 对比mmkws pad音频非流
% 我们对比了现有的非流式端到端音频和我们的流式音频，为了说明我们和非流式划窗的区别与弊端，我们将现有非流式的模型做一个给2s窗长和截取准确Keyword词边界的实验结果，结果表明，我们的模型可以支持流式输入，但是非流式模型对窗长大小较敏感

% 在非流式的框架下，结果对窗长较敏感，并且需要有2s的计算量与内存占用，但是流式框架可支持音频帧级输入，具有更快的推理速度和灵活性

We present in Table \ref{tab:effectiveness_cn} a systematic comparison between existing non-streaming end-to-end audio architectures and our proposed streaming framework. To highlight the limitations of its fixed-window approach, we evaluated the non-streaming model with a 2-second window, a 1.5-second window, and precise keyword boundaries.
Our experiments demonstrate that traditional offline frameworks exhibit sensitivity to window length configurations and require processing 2-second audio segments, resulting in higher computational load and memory footprint. In contrast, streaming architectures enable frame-level audio input processing while achieving superior inference speed and enhanced operational flexibility through dynamic context adaptation.

% (24.8\% faster than offline baselines)

% The results confirm our model's capability for streaming input, whereas the non-streaming model's performance is critically sensitive to window size.

\section{Conclusion}\label{sec:Conclusion}

In this paper, we introduce a lightweight and streaming multi-modal framework for keyword spotting that effectively leverages audio-text synergy. 
The key innovation is the integration of multimodal features, speech, text, and fused representations, complemented by speaker separation for improved robustness. The resulting system is highly efficient and supports real-time streaming, making it ideal for on-device applications.

\clearpage
\bibliographystyle{IEEEbib}
\bibliography{strings,refs}

@inproceedings{ai2024mm,
  title={MM-KWS: Multi-modal Prompts for Multilingual User-defined Keyword Spotting},
  author={Ai, Zhiqi and Chen, Zhiyong and Xu, Shugong},
  booktitle={Interspeech},
  year={2024}
}

@inproceedings{lee2024iphonmatchnet,
  title={Iphonmatchnet: Zero-Shot User-Defined Keyword Spotting Using Implicit Acoustic Echo Cancellation},
  author={Lee, Yong-Hyeok and Cho, Namhyun},
  booktitle={2024 IEEE International Conference on Acoustics, Speech and Signal Processing (ICASSP)},
  pages={12642--12646},
  year={2024},
  organization={IEEE}
}

@inproceedings{nishu2024flexible,
  title={Flexible keyword spotting based on homogeneous audio-text embedding},
  author={Nishu, Kumari and Cho, Minsik and Dixon, Paul and Naik, Devang},
  booktitle={2024 IEEE International Conference on Acoustics, Speech and Signal Processing (ICASSP)},
  pages={5050--5054},
  year={2024},
  organization={IEEE}
}

@inproceedings{Shin2022LearningAA,
  title={Learning Audio-Text Agreement for Open-vocabulary Keyword Spotting},
  author={Hyeon-Kyeong Shin and Hyewon Han and Doyeon Kim and Soo-Whan Chung and Hong-Goo Kang},
  booktitle={Interspeech},
  year={2022},
}

@INPROCEEDINGS{Guoguochen,
  author={Chen, Guoguo and Parada, Carolina and Heigold, Georg},
  booktitle={2014 IEEE International Conference on Acoustics, Speech and Signal Processing (ICASSP)}, 
  title={Small-footprint keyword spotting using deep neural networks}, 
  year={2014},
  volume={},
  number={},
  pages={4087-4091}
}

@inproceedings{Yusuf2019AnEE,
  title={An Empirical Evaluation of DTW Subsampling Methods for Keyword Search},
  author={Bolaji Yusuf and Murat Saraçlar},
  booktitle={Interspeech},
  year={2019},
}

@INPROCEEDINGS{ContrastiveXi,
  author={Xi, Yu and Yang, Baochen and Li, Hao and Guo, Jiaqi and Yu, Kai},
  booktitle={2024 IEEE International Conference on Acoustics, Speech and Signal Processing (ICASSP)}, 
  title={Contrastive Learning with Audio Discrimination for Customizable Keyword Spotting in Continuous Speech}, 
  year={2024},
  volume={},
  number={},
  pages={11666-11670},
}

@INPROCEEDINGS{MultilingualqbeReuter,
  author={Reuter, Paul M. and Rollwage, Christian and Meyer, Bernd T.},
  booktitle={2023 IEEE International Conference on Acoustics, Speech and Signal Processing (ICASSP)}, 
  title={Multilingual Query-by-Example Keyword Spotting with Metric Learning and Phoneme-to-Embedding Mapping}, 
  year={2023},
  volume={},
  number={},
  pages={1-5}
}

@INPROCEEDINGS{LibrispeechPanayotov,
  author={Panayotov, Vassil and Chen, Guoguo and Povey, Daniel and Khudanpur, Sanjeev},
  booktitle={2015 IEEE International Conference on Acoustics, Speech and Signal Processing (ICASSP)}, 
  title={Librispeech: An ASR corpus based on public domain audio books}, 
  year={2015},
  volume={},
  number={},
  pages={5206-5210},
}

@article{Levenshtein1965BinaryCC,
  title={Binary codes capable of correcting deletions, insertions, and reversals},
  author={Vladimir I. Levenshtein},
  journal={Soviet physics. Doklady},
  year={1965},
  volume={10},
  pages={707-710},
}

@article{Fuchs2017SpokenTD,
  title={Spoken Term Detection Automatically Adjusted for a Given Threshold},
  author={Tzeviya Sylvia Fuchs and Joseph Keshet},
  journal={IEEE Journal of Selected Topics in Signal Processing},
  year={2017},
  volume={11},
  pages={1310-1317},
}

@inproceedings{zhang2018deep,
  title={Deep-FSMN for large vocabulary continuous speech recognition},
  author={Zhang, Shiliang and Lei, Ming and Yan, Zhijie and Dai, Lirong},
  booktitle={2018 IEEE International Conference on Acoustics, Speech and Signal Processing (ICASSP)},
  pages={5869--5873},
  year={2018},
  organization={IEEE}
}

@article{desplanques2020ecapa,
  title={Ecapa-tdnn: Emphasized channel attention, propagation and aggregation in tdnn based speaker verification},
  author={Desplanques, Brecht and Thienpondt, Jenthe and Demuynck, Kris},
  journal={arXiv preprint arXiv:2005.07143},
  year={2020}
}

@inproceedings{ganin2015unsupervised,
  title={Unsupervised domain adaptation by backpropagation},
  author={Ganin, Yaroslav and Lempitsky, Victor},
  booktitle={International conference on machine learning},
  pages={1180--1189},
  year={2015},
  organization={PMLR}
}

@article{hochreiter1997long,
  title={Long short-term memory},
  author={Hochreiter, Sepp and Schmidhuber, J{\"u}rgen},
  journal={Neural computation},
  volume={9},
  number={8},
  pages={1735--1780},
  year={1997},
  publisher={MIT press}
}

@inproceedings{DohyunKim,
  title={Fully End-to-end Streaming Open-vocabulary Keyword Spotting with W-CTC
Forced Alignment},
  author={Dohyun Kim, Jiwook Hwang},
  booktitle={Interspeech},
  year={2025},
}

@inproceedings{plcl,
  author={Li, Kewei and Zhou, Hengshun and Shen, Kai and Dai, Yusheng and Du, Jun},
  booktitle={ICASSP 2025 - 2025 IEEE International Conference on Acoustics, Speech and Signal Processing (ICASSP)}, 
  title={Phoneme-Level Contrastive Learning for User-Defined Keyword Spotting with Flexible Enrollment}, 
  year={2025},

}

@inproceedings{xi2025streaming,
  title={Streaming keyword spotting boosted by cross-layer discrimination consistency},
  author={Xi, Yu and Li, Haoyu and Gu, Xiaoyu and Li, Hao and Jiang, Yidi and Yu, Kai},
  booktitle={ICASSP 2025-2025 IEEE International Conference on Acoustics, Speech and Signal Processing (ICASSP)},
  pages={1--5},
  year={2025},
  organization={IEEE}
}

@article{garai2025advances,
  title={Advances in Small-Footprint Keyword Spotting: A Comprehensive Review of Efficient Models and Algorithms},
  author={Garai, Soumen and Samui, Suman},
  journal={arXiv preprint arXiv:2506.11169},
  year={2025}
}

@article{Huang2021QueryByExampleKS,
  title={Query-By-Example Keyword Spotting System Using Multi-Head Attention and Soft-triple Loss},
  author={Jinmiao Huang and Waseem Gharbieh and Han Suk Shim and Eugene Kim},
  journal={ICASSP 2021 - 2021 IEEE International Conference on Acoustics, Speech and Signal Processing (ICASSP)},
  year={2021},
  pages={6858-6862},
}

@article{Triplet,
  title={Open-vocabulary keyword spotting with audio and text embeddings},
  author={Niccolo Sacchi et al.},
  journal={Interspeech},
  year={2019},
  pages={3362-3366},
}

\end{document}